\documentclass[12pt]{article}

\newcommand{\x}{arXiv:}
\newcommand{\m}{\mathrm}
\newcommand{\be}{\begin{equation}}
\newcommand{\ee}{\end{equation}}
\newcommand{\ba}{\begin{eqnarray}}
\newcommand{\ea}{\end{eqnarray}}

\usepackage{graphicx}
\usepackage{amssymb}
\usepackage{amsmath}
\usepackage[T1]{fontenc} 
\usepackage[ansinew]{inputenc} 
\usepackage[nosort]{cite}
\newcommand{\inbar}{\vrule height1.57ex width.4pt depth0pt}
\newcommand{\SW}{\relax{\hbox{$\ \inbar\kern-.285em{\rm S}$}}}

\begin{document}
\thispagestyle{empty}
\begin{center}

\null \vskip-1truecm \vskip2truecm

{\Large{\bf \textsf{Holography of Low-Centrality Heavy Ion Collisions}}}
\vskip0.5truecm
{\Large{\bf \textsf{}}}

\vskip0.5truecm
{\Large{\bf \textsf{}}}

\vskip1truecm

{\large \textsf{Brett McInnes
}}

\vskip0.1truecm

\textsf{\\ National
  University of Singapore}
  \vskip1.2truecm
\textsf{email: matmcinn@nus.edu.sg}\\

\end{center}
\vskip1truecm \centerline{\textsf{ABSTRACT}} \baselineskip=15pt
\medskip

Large vorticities in the Quark-Gluon Plasma produced in peripheral collisions studied by the STAR collaboration at the RHIC facility have been deduced from observations of polarizations of $\Lambda$ and $\overline{\Lambda}$ hyperons. Recently, the STAR collaboration has reported on the dependence of these polarizations on centrality, at impact energy 200 GeV and relatively large centralities $\mathcal{C} \geq 20\%$. The polarizations increase slowly with centrality, up to perhaps $\mathcal{C} = \,60 - 70\%$. Here we use a holographic model of the vortical QGP to make predictions regarding these polarizations for smaller centralities, ranging from $5 - 20\%$. The model predicts that as one moves downwards from $20\%$,  $\Lambda/\overline{\Lambda}$ polarizations at first decrease but then \emph{increase} sharply, in a characteristic pattern which should be readily detected if collisions can be studied at impact energies below 200 GeV and centrality as low as $5 - 10\%$. The effect should be most evident at moderate impact energies below 200 GeV, so we give predictions for impact energy 27 GeV.

\newpage
\addtocounter{section}{1}
\section* {\large{\textsf{1. The Vorticity of the Quark-Gluon Plasma}}}
It has long been predicted \cite{kn:liang,kn:bec,kn:huang} that peripheral collisions of heavy ions should, for impact energies high enough to produce a Quark-Gluon Plasma (QGP), give rise to large \emph{vorticities} in that plasma, corresponding to local angular velocities on the order of $10^{21}$ -- $10^{22}\,\cdot\,$s$^{-1}$. Fortunately, there is a particular observable, the global polarization of $\Lambda$ and $\overline{\Lambda}$ hyperons \cite{kn:hyper}, which was expected to manifest this phenomenon in a relatively straightforward way, and the STAR collaboration \cite{kn:STAR} at the RHIC facility has succeeded \cite{kn:STARcoll,kn:STARcoll2,kn:STARcoll3,kn:STARcoll4} in observing it\footnote{Note that this polarization is just one aspect of the angular momentum imparted to the plasma (and is associated with only a small fraction of it): hyperon polarization is simply the aspect which is easiest to detect. Furthermore, doubts have been raised (see for example \cite{kn:M1,kn:M2}) regarding the nature of the relationship between hyperon polarization and QGP vorticity. While one should be aware of these issues, here we simply follow \cite{kn:STARcoll}.}.

The vorticity imparted to the QGP in a peripheral collision depends primarily on two parameters, the impact energy and the centrality. Data on the polarization of $\Lambda$ and $\overline{\Lambda}$ hyperons as a (rapidly decreasing) function of impact energy, at essentially fixed centrality, were given in both \cite{kn:STARcoll} and \cite{kn:STARcoll2}. Data on this polarization as a (slowly increasing) function of centrality, at fixed impact energy 200 GeV, were also given in \cite{kn:STARcoll2}, though only at relatively large centralities ($\geq 20\%$). \emph{Whether the pattern observed there persists at lower centralities remains an open question}\footnote{Searches for $\Lambda/\overline{\Lambda}$ hyperon polarization have been conducted \cite{kn:bed} at such low centralities at the ALICE detector at the LHC. Unfortunately, at the high impact energies produced by the LHC, the polarizations are (at present) too small to be detected. (Our model, described below, predicts for these collisions a total $\Lambda + \overline{\Lambda}$ polarization in the range $0.046 - 0.055\%$, in agreement with recent hydrodynamic calculations \cite{kn:becky}.)}.

In an earlier work \cite{kn:93}, briefly summarized in the next section, we constructed a simple \emph{holographic} model \cite{kn:nat} of the ``vortical QGP'', in which there is an \emph{upper bound} on the possible vorticities of fluids dual to rotating five-dimensional asymptotically AdS black holes with given specific angular momenta\footnote{See \cite{kn:sonner,kn:schalm} for earlier holographic applications of \emph{four}-dimensional AdS-Kerr black holes.}. The bound takes the form
\begin{equation}\label{ALPHA}
\omega\;\leq\;\varkappa\;{\varepsilon\over \alpha}\;\approx \; 0.2782\;{\varepsilon\over \alpha}\,,
\end{equation}
where $\omega$ is the vorticity of the dual fluid on the boundary, $\alpha$ is its angular momentum density, $\varepsilon$ is its energy density, and $\varkappa\,$  is a dimensionless constant with the indicated approximate value; we briefly review this in the next section. We showed that this upper bound is in fact in surprisingly good agreement with the STAR data on $\Lambda/\overline{\Lambda}$ hyperon polarization as a function of impact energy: that is, the data suggest strongly that this upper bound is actually saturated, and we adopt this saturation of (\ref{ALPHA}) (under conditions we will discuss later) as our working hypothesis. Notice in particular that this hypothesis suggests that vorticity decreases with impact energy (since $\alpha$ increases with impact energy much more rapidly than $\varepsilon$), in agreement with the data \cite{kn:STARcoll,kn:STARcoll2}.

The holographic model focuses our attention on a key quantity, $\alpha/\varepsilon$. This is almost inevitable, for, in the theory of rotating black holes generally, the ratio of the black hole's angular momentum to its mass is universally taken (for example, in astrophysical studies of actual black holes) to be a central object of interest; and the holographic dual of this quantity in the asymptotically AdS case is indeed $\alpha/\varepsilon$.

This quantity has several virtues in our case: for example, as a ratio of densities, it is much more stable as the plasma expands than either $\alpha$ or $\varepsilon$ alone; and it can be straightforwardly computed from the impact energy and the centrality of the collision\footnote{It also clarifies other issues. For example, notice carefully that, throughout this work, the angular momentum always enters \emph{only} through its density, $\alpha$: the total angular momentum never appears. Thus for example the fact that hyperon polarization only represents a small fraction of the total angular momentum of the full system is not relevant.}. Another virtue of $\alpha/\varepsilon$ is that it can be computed from phenomenological models for collisions with very small centralities (below $20\%$), thus allowing us to complement the experimental data for larger centralities, as follows.

As mentioned above, the new data analysis reported in \cite{kn:STARcoll2} yields information on the $\Lambda/\overline{\Lambda}$ polarizations as a function of centrality, for collisions at $\sqrt{s_{\m{NN}}} = 200$ GeV and for centralities $\geq 20\%$. It is found that the  polarizations grow slowly with centrality, apparently reaching a maximum at $\mathcal{C} = 60 - 70\%$. One might expect to find that, if one could move downward from $\mathcal{C} = 20\%$, the polarizations would become steadily smaller, continuing the observed trend.

Reliable phenomenological models \cite{kn:jiang} compute the angular momentum in peripheral collisions with small impact parameters (or centralities \cite{kn:bron}), and these can be used to compute the angular momentum \emph{densities} at small centralities; similarly there are phenomenological estimates \cite{kn:sahoo} of the energy densities corresponding to central collisions at various impact energies. Then (\ref{ALPHA}) allows us to put upper bounds on vorticity at low centralities, assuming that $\varepsilon$ has much the same value in low-centrality collisions as in central collisions. Assuming that the bound continues to be saturated, we can predict or at least roughly estimate the shape of the polarization vs. centrality curve at these low centralities.

In this work, we carry out this programme, giving predictions as to what might be found if lower centralities, down to $5\%$, can be investigated in the beam energy scan experiments. We find that the model predicts, as expected, an initial continued fall in the polarizations as $\mathcal{C}$ drops below $20\%$, but that there is a global minimum at around $\mathcal{C} = 17\%$; the polarizations steadily \emph{increase} for still smaller centralities, becoming quite marked in the centrality range $5 - 10\%$. As discussed above, polarizations tend to be more prominent at lower impact energies, say for example $\sqrt{s_{\m{NN}}} = 27$ GeV, and should appear clearly in that case; we give detailed predictions for this.

\addtocounter{section}{1}
\section* {\large{\textsf{2. Brief Review of the Vorticity Bound}}}
The vorticity bound is an application of the gauge-gravity duality \cite{kn:nat}, which (in the current application) is an equivalence between the physics of a five-dimensional AdS-like spacetime and that of an $\mathcal{N} = 4$ super-Yang-Mills field theory on the four-dimensional conformal boundary. It must be admitted from the outset that this field theory is not very much like Quantum Chromodynamics, at least not in all cases. In specific, concrete applications, however, there are arguments that such theories are not as different from QCD as they first appear: see \cite{kn:bena,kn:endrodi} for two recent very remarkable examples. One can be cautiously optimistic that, in the domain of temperatures and baryonic chemical potentials in which both these boundary theories and QCD are strongly coupled (which is likely to be the case for the plasmas produced in the beam energy scans, which are our primary concern), there may be \emph{universal} behaviour (see Section 12.2.3 of \cite{kn:nat}) which permits lessons learned in the former to be applied to the latter.

The relevant bulk spacetime here is the five-dimensional AdS-Kerr black hole spacetime rotating about one of the two possible axes: see \cite{kn:hawk,kn:cognola,kn:gibperry}. The metric is
\begin{flalign}\label{AROD}
g(\m{AdSK}_5^{(a,0)}) = &- {\Delta_r \over \rho^2}\Bigg[\,\m{d}t \; - \; {a \over \Xi}\m{sin}^2\theta \,\m{d}\phi\Bigg]^2\;+\;{\rho^2 \over \Delta_r}\m{d}r^2\;+\;{\rho^2 \over \Delta_{\theta}}\m{d}\theta^2 \\ \notag \,\,\,\,&+\;{\m{sin}^2\theta \,\Delta_{\theta} \over \rho^2}\Bigg[a\,\m{d}t \; - \;{r^2\,+\,a^2 \over \Xi}\,\m{d}\phi\Bigg]^2 \;+\;r^2\cos^2\theta \,\m{d}\psi^2 ,
\end{flalign}
where
\begin{eqnarray}\label{BROD}
\rho^2& = & r^2\;+\;a^2\m{cos}^2\theta, \nonumber\\
\Delta_r & = & (r^2+a^2)\left(1 + {r^2\over L^2}\right) - 2M,\nonumber\\
\Delta_{\theta}& = & 1 - {a^2\over L^2} \, \m{cos}^2\theta, \nonumber\\
\Xi & = & 1 - {a^2\over L^2}.
\end{eqnarray}
Here ($\theta,\,\phi,\,\psi$) are Hopf coordinates on the three-dimensional sphere, $L$ is the asymptotic AdS curvature length scale, and $M$ and $a$ are parameters describing the geometry of the spacetime. Clearly $a$ controls the rate of rotation; it is a (non-trivial) function of $L$ and of the angular momentum per unit mass of the black hole.

One can show (see \cite{kn:93} for the details) that the ``holographic dictionary'' in this case is as follows.

As always, the Hawking temperature of the black hole is identified with the temperature of the plasma-like matter in the boundary theory, $T_{\infty}$:
\begin{equation}\label{EROD}
T_{\infty}\;=\;{r_H\left(1 + {r_H^2\over L^2}\right)\over 2\pi \left(r_H^2 + a^2\right)} + {r_H\over 2\pi L^2}.
\end{equation}
The entropy per unit physical mass of the black hole is holographically identified with the ratio, $s/\varepsilon$, of the entropy density to the energy density of the boundary matter:
\begin{equation}\label{FROD}
{s\over \varepsilon}\;=\;{2\pi r_H\left(r_H^2+a^2\right)\left(1 - {a^2\over L^2}\right)\over M\left(3 - {a^2\over L^2}\right)}.
\end{equation}
In \cite{kn:93}, we considered a plasma on the boundary with a specified ratio $\mathcal{A}$ of angular momentum to energy densities, representing it by particles with an angular momentum to mass ratio $\mathcal{A}$ and an angular velocity (relative to a frame co-rotating with the boundary) $\omega$. The dual system is naturally assumed to be the black hole we have been discussing, with angular momentum per unit mass \emph{also} equal to $\mathcal{A}$. This assumption means that the value of $\mathcal{A}$ affects the bulk geometry. But the holographic duality implies that the bulk geometry determines the boundary geometry. In this way we obtain a holographic relation between $\omega$ and $\mathcal{A}$ (and $L$). One finds that
\begin{equation}\label{GROD}
\omega\;=\;{\mathcal{A}\over L^2}\,\sqrt{{\Xi\over 1\;+\;{\mathcal{A}^2\over L^2}\,\Xi}},
\end{equation}
where $\Xi$ is defined in (\ref{BROD}). (Note that $\Xi$ is a function of $a$ and $L$, which means (see above) that it can be regarded as a function\footnote{Explicitly, one has $
\Xi\;=\;{\sqrt{1\,+\,{3\mathcal{A}^2\over L^2}}\;-\;1\;-\;{\mathcal{A}^2\over L^2}\over {\mathcal{A}^2\over 2L^2}}$.} of $\mathcal{A}$ and $L$.)

A straightforward calculation shows (see again \cite{kn:93}) that, if $\mathcal{A}$ is fixed, as above, by the impact energy and centrality of a given collision, then $\omega$, regarded formally as a function of $L$, is bounded:
\begin{equation}\label{HROD}
\omega\;\leq\;\varkappa\;{\varepsilon\over \alpha}\,,
\end{equation}
where $\varkappa$ is a constant, $\varkappa\;\approx \; 0.2782$. This is the vorticity bound above, the inequality (\ref{ALPHA}). The bound is saturated when $\mathcal{A}$ is such that $L/\mathcal{A} \approx 1.2499$.

This bound can be translated to a bound on the total polarization of $\Lambda/\overline{\Lambda}$ hyperons produced in heavy ion collisions, as follows. As above, we use phenomenological models \cite{kn:jiang,kn:sahoo} to obtain values for $\varepsilon$, $\alpha$, and also the temperature $T$, for collisions at given impact energies and centralities. Using the formula \cite{kn:hyper,kn:STARcoll} for the total average polarization,
\begin{equation}\label{F}
\overline{\mathcal{P}}_{\Lambda'}\,+\,\overline{\mathcal{P}}_{\overline{\Lambda}'}\;=\;{\omega\over T},
\end{equation}
we can compute an upper bound on this total polarization. (The primes refer to the fact that this applies to primary hyperons, see \cite{kn:hyper}.) That is, we obtain inequalities of the form
\begin{equation}\label{G}
\left[\overline{\mathcal{P}}_{\Lambda'}\,+\,\overline{\mathcal{P}}_{\overline{\Lambda}'}\right]\left(\sqrt{s_{\m{NN}}},\,\mathcal{C}\right) \;\leq \; \Phi(\sqrt{s_{\m{NN}}}, \mathcal{C})\;\equiv \;{\varkappa\,\varepsilon \over \alpha \,T},
\end{equation}
where $\mathcal{C}$ denotes centrality, and where all of the quantities on the right are (approximately) known to us, for each impact energy and centrality.
\addtocounter{section}{1}
\section* {\large{\textsf{3. Total Polarization vs. Impact Energy}}}
We have carried out this programme in such a manner as to enable a direct comparison with the data reported in \cite{kn:STARcoll,kn:STARcoll2} regarding the variation of observed hyperon polarizations with impact energy.
\begin{figure}[!h]
\centering
\includegraphics[width=1\textwidth]{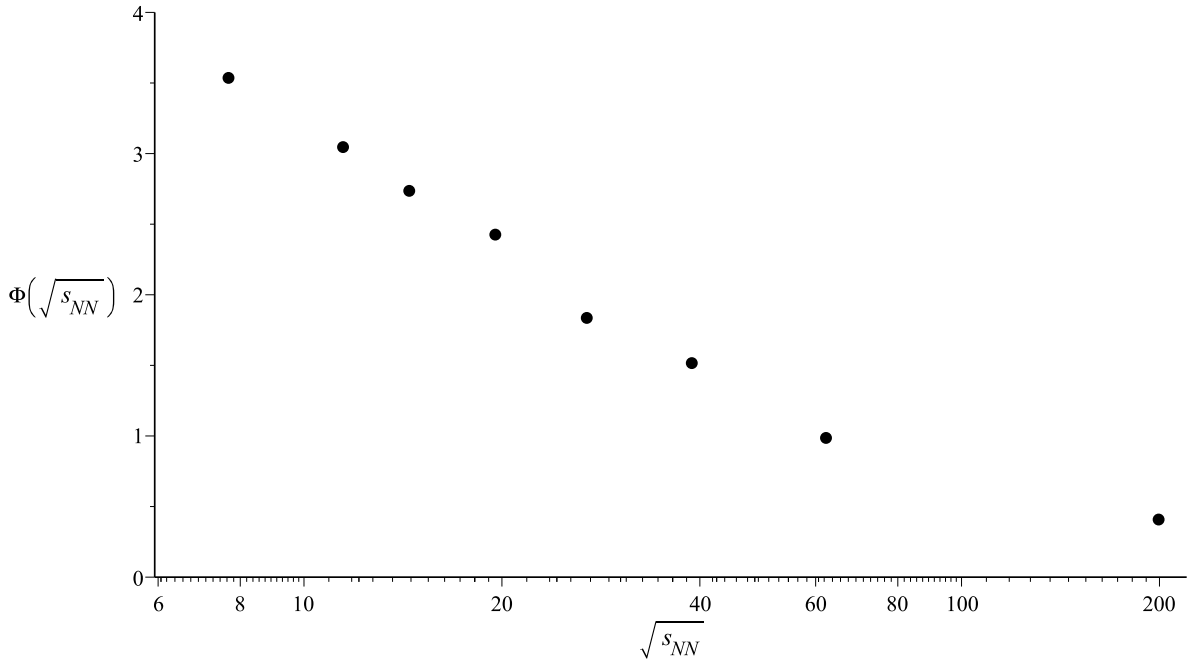}
\caption{Theoretical upper bounds on total polarization, that is, $ \left[\overline{\mathcal{P}}_{\Lambda'}\,+\,\overline{\mathcal{P}}_{\overline{\Lambda}'}\right]\left(\sqrt{s_{\m{NN}}},\,\mathcal{C} = 20\%\right) \,\leq \,  \Phi\left(\sqrt{s_{\m{NN}}},\, \mathcal{C} = 20\%\right)$, as a percentage, for collisions at $\sqrt{s_{\m{NN}}} = 7.7,\, 11.5,\, 14.5,\, 19.6,\, 27,\, 39,\, 62.4,\, 200$ GeV and $20\%$ centrality.}
\end{figure}
The results, for collisions at centrality\footnote{References \cite{kn:STARcoll,kn:STARcoll2} report averaged data for collisions with $\mathcal{C} = 20 - 50\%$. The variation in the total polarization in this specific range is relatively very slow, as can be seen from Figure 5 in \cite{kn:STARcoll2}, so we focus on $\mathcal{C} = 20\%$ (impact parameter around 6.75 femtometres) to ensure that the results are not affected by edge effects or an excessively small overlap volume.} $\mathcal{C} = 20\%$, are summarized in Figure 1 (taken from \cite{kn:93}), which is to be compared with Figure 4 of \cite{kn:STARcoll} and Figure 4 of \cite{kn:STARcoll2} by adding together, in the latter, the values corresponding to the two points (corresponding to $\Lambda$ and $\overline{\Lambda}$ hyperons) at each impact energy.

We see from this comparison that the vorticity bound is, to a good approximation, respected: indeed, it is approximately \emph{attained}, at least for collisions at impact energies between 11.5 and 200 GeV.

Under what circumstances should we expect the bound to be attained? Clearly, since the bound is inversely proportional to $\alpha$, which is nearly zero for extremely low-centrality collisions, we must not expect it to be attained under such conditions. Similar comments apply to extremely high-centrality collisions. Thus, we should confine ourselves to collisions in a definite range of impact parameters.

As mentioned, the bound is satisfied (and is attained) for collisions at $\sqrt{s_{\m{NN}}} = 11.5$ GeV and centrality $20\%$. In such a collision, the angular momentum density is $\alpha \approx 44$ fm$^{-3}$. Let us assume that the question as to whether the bound is attained is determined by $\alpha$. In view of the computation just described, we will not assume that the bound is attained at any angular momentum density below around $40$ fm$^{-3}$. For collisions at 19.6 GeV, the lowest impact energy for which one can be entirely confident that a plasma forms, and at which the bound is clearly attained, this translates to an impact parameter no lower than 2.5 fm (or centrality about 2.5$\%$).

On the other hand, in this work we are concerned with centralities up to $20\%$, corresponding to impact parameter around 6.75 fm; so we arbitrarily set 7 fm as a (very conservative) upper limit beyond which we do not insist that the vorticity bound is attained. In short, for the collisions of interest in the present work, that is, with impact energies of at least 19.6 GeV and centralities no larger than $20\%$, the range of impact parameters with which we are concerned is $2.5 - 7$ fm.

We see, then, that the holographic model is in good agreement with the existing data on QGP vorticity as a function of impact energy. Let us now discuss a regime in which data have yet to be reported, and try to use this model to make some predictions as to what will be seen when they are.

\addtocounter{section}{1}
\section* {\large{\textsf{4. Total Polarization vs. Centrality: Predictions for $\mathcal{C} < 20\%$ }}}
In \cite{kn:STARcoll2}, the STAR collaboration reports data on $\Lambda$ and $\overline{\Lambda}$ hyperon polarization in a novel way: as a function of centrality (for fixed impact energy $\sqrt{s_{\m{NN}}} = 200$ GeV), instead of, as above, the other way around. The uncertainties are somewhat large at present, but overall patterns are discernible. In particular, the polarizations apparently increase slowly with centrality (Figure 5 of \cite{kn:STARcoll2}), from $\mathcal{C} = 20\%$ upwards, perhaps as far as $\mathcal{C} = 60 - 70\%$; beyond that, there is some suggestion of a decrease, though the statistics do not permit a definite conclusion.

We now ask: what happens if collisions can be studied at centralities \emph{below} $20\%$?

There are (at least) two competing factors here. As the centrality decreases from $20\%$, the angular momentum imparted to the plasma increases dramatically (see Figure 3 of \cite{kn:jiang}), tending to increase the vorticity. On the other hand, the moment of inertia might be expected ---$\,$ though, in view of the complex relation between the moment of inertia and the dynamics here, the extent of this effect is far from clear ---$\,$ to increase, which would lower the vorticity. Predictions as to the vorticities and polarizations to be expected in this regime depend on which of the two competing effects one expects to dominate: they are model-dependent. One can anticipate that, moving down from $\mathcal{C} = 20\%$, the pattern observed at high centralities should continue, that is, the total polarization should drop at first. The question is whether this decline continues for very low centralities ---$\,$ meaning for extremely high angular momenta, high even compared to those in plasmas for which data have been reported.

The holographic model incorporates all of these factors in a concrete way, and answers our questions explicitly. It can be used, in the same way as above, to generate upper bounds for the total average polarization as centrality varies. In order to understand the results, one should bear in mind the predicted shape of the function (again, Figure 3 of \cite{kn:jiang}) giving angular momentum as a function of impact parameter $b$. It is of course zero for perfectly central collisions ($b = 0$), but then it rises rapidly to reach a maximum around $b = 4$ fm. It then decreases monotonically and rapidly (and almost linearly) for all larger $b$.

However, the angular momentum \emph{density} behaves in a somewhat more complicated way, because the volume by which one is dividing decreases steadily with $b$. The upshot is that \emph{it is not clear that $\alpha$ will be large at low centralities, even though the angular momentum itself may be very large in such cases}\footnote{We assume that, for low centralities, the quantity $\varepsilon/T$ on the right side of equation (\ref{G}) is approximately constant as a function of centrality. This may not be a good approximation for large centralities, and in fact the variations of these parameters at very high centralities are not well-understood (see \cite{kn:quench,kn:quenchagain,kn:quenchyetagain}). For this reason we do not (yet) attempt to use our model for very large centralities. Of course, one can consider centralities a little beyond $20\%$, and in that range the model does suggest a slow increase of polarization with centrality (see Figures 2 and 3, below), in agreement with \cite{kn:STARcoll2}.}. Thus, one cannot foresee how the crucial quantity $\kappa\varepsilon/\alpha$ will behave at low centrality: one has to compute.

The computed upper bounds, for $\sqrt{s_{\m{NN}}} = $ 200 GeV, are as shown in Figure 2, for centralities ranging from $5\%$ to $20\%$. \emph{If} our hypothesis that the vorticity bound is attained in collisions with impact energy at least 19.6 GeV and a range of impact parameters between 2.5 and 7 fm (which encompasses $5\%$ to $20\%$ centrality) is correct, then these are predictions for the total average polarization in such collisions. (Of course, it may be that the bound is correct but this additional hypothesis is not; this would mean that the bound is attained in some collisions but not in others with almost identical angular momentum densities, which would be very puzzling.)
\begin{figure}[!h]
\centering
\includegraphics[width=1.1\textwidth]{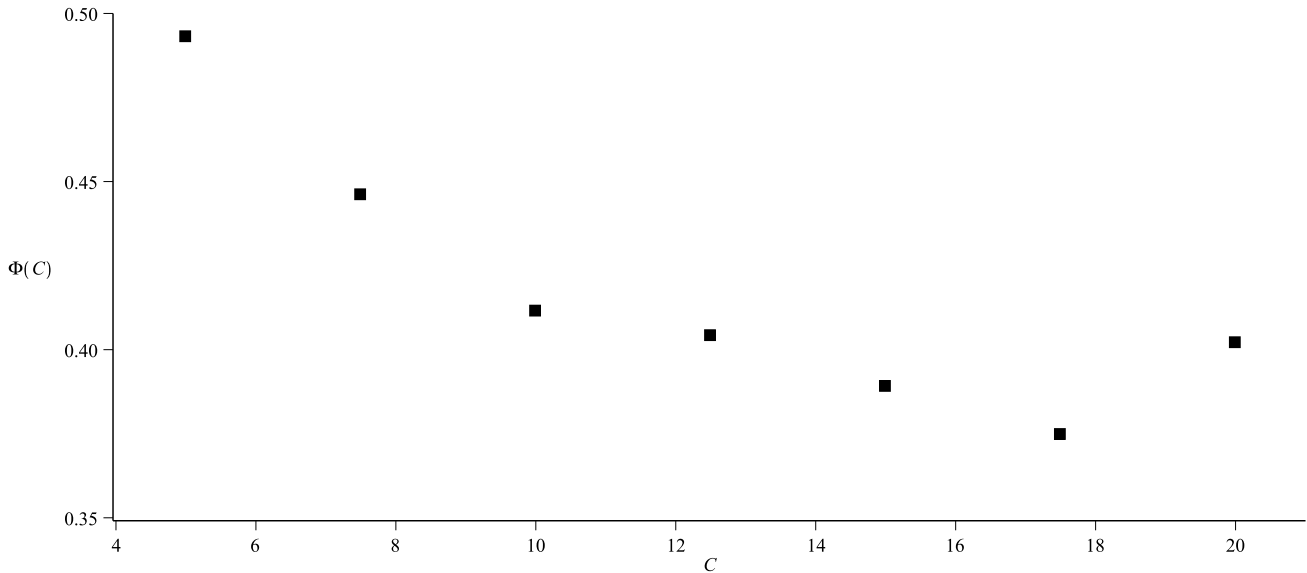}
\caption{Theoretical upper bounds on total polarization, that is, $ \left[\overline{\mathcal{P}}_{\Lambda'}\,+\,\overline{\mathcal{P}}_{\overline{\Lambda}'}\right]\left(\sqrt{s_{\m{NN}}} = 200 \; \m{GeV},\,\mathcal{C}\right) \leq \Phi\left(\sqrt{s_{\m{NN}}} = 200 \; \m{GeV},\, \mathcal{C}\right)$, as a percentage, for collisions with centrality $\mathcal{C}$ ranging from $5\%$ to $20\%$ and impact energy 200 GeV.}
\end{figure}
We see that, as one moves down from $\mathcal{C} = 20\%$, the predicted bound drops at first, as might be expected from the STAR data at higher centralities; soon, however, \emph{the bound begins to rise} slowly, regaining the value at $\mathcal{C} = 20\%$ at around $12.5\%$, and then rising steadily more rapidly for smaller centralities.

Earlier investigations by the STAR collaboration showed, as mentioned above, that the polarizations are much larger at impact energies lower than 200 GeV; and in fact the evidence for polarization is most unambiguous for $\sqrt{s_{\m{NN}}} =  19.6,\, 27,\, 39$ GeV. The calculations leading to Figure 2 can readily be repeated in these cases: Figure 3 shows the results for $\sqrt{s_{\m{NN}}} =  27$ GeV.
\begin{figure}[!h]
\centering
\includegraphics[width=1.1\textwidth]{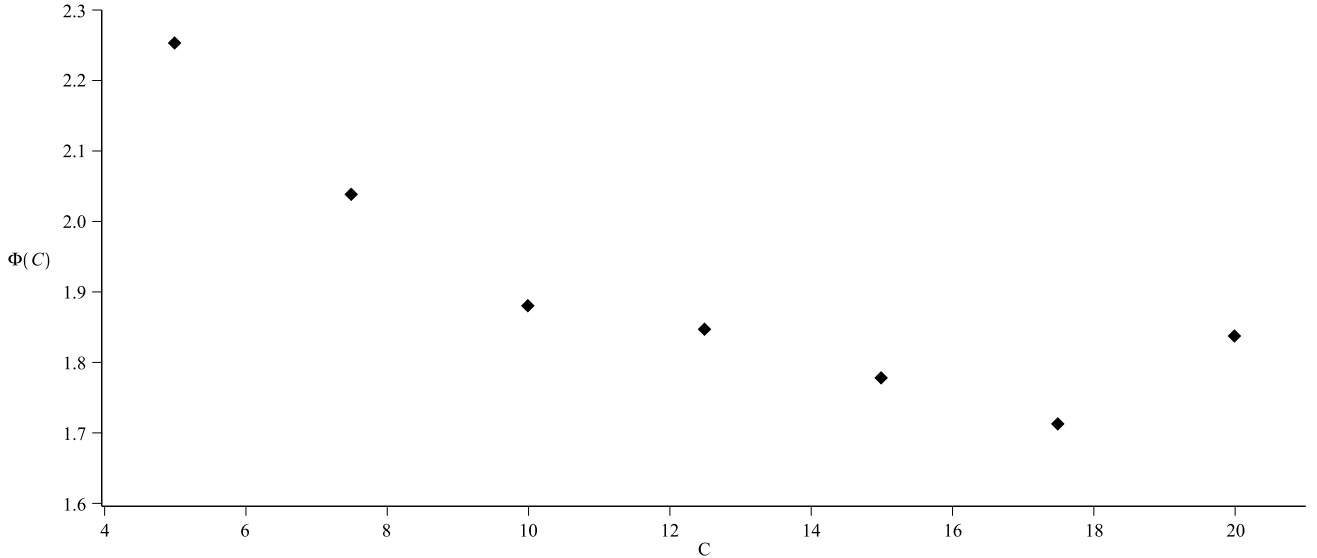}
\caption{Theoretical upper bounds on total polarization, that is, $ \left[\overline{\mathcal{P}}_{\Lambda'}\,+\,\overline{\mathcal{P}}_{\overline{\Lambda}'}\right]\left(\sqrt{s_{\m{NN}}} = 27 \; \m{GeV},\,\mathcal{C}\right) \leq \Phi\left(\sqrt{s_{\m{NN}}} = 27\; \m{GeV},\, \mathcal{C}\right)$, as a percentage, for collisions with centrality $\mathcal{C}$ ranging from $5\%$ to $20\%$ and impact energy 27 GeV.}
\end{figure}
These are the predictions of the holographic model in this case: the overall shape of the graph is as in Figure 2, in fact it is characteristic of this model; but the vertical scale is substantially larger. Given that polarization is clearly seen at this impact energy and centrality $\mathcal{C} = 20\%$, we can predict that it would be at least as clearly discernible in collisions at this impact energy and $\mathcal{C} = 5 - 10\%$. To be specific: at $\mathcal{C} = 7.5\%$, we expect the total polarization to be about $11\%$ larger than in collisions with $\mathcal{C} = 20\%$; at $\mathcal{C} = 5\%$, we expect it to be about $23\%$ larger.

\section* {\large{\textsf{5. Conclusion}}}
The polarizations of $\Lambda$ and $\overline{\Lambda}$ hyperons produced in heavy ion collisions have been observed \cite{kn:STARcoll,kn:STARcoll2} to behave in a consistent way with respect to variations of the impact energy: they decrease as the impact energy increases. We have seen (Figure 1) that the holographic model is compatible with this observation.

The STAR data also suggest \cite{kn:STARcoll2} a new avenue for exploration: the effect of varying \emph{centrality} on $\Lambda$ and $\overline{\Lambda}$ hyperon polarizations, at fixed impact energy. In this case, data have yet to be reported for centralities lower than $20\%$  with impact energies at 200 GeV and below. The holographic vorticity bound, supplemented by the (observationally founded) conjecture that the bound is approximately attained in collisions with impact energies at least 19.6 GeV and impact parameters in the $2.5 - 7$ femtometre range, \emph{predicts definitely larger total polarizations, perhaps by as much as $10 - 20\%$}, in low-centrality collisions as compared to those with $\mathcal{C} = 20\%$. It will be interesting to see how these predictions fare, if data can be taken at this impact energy and in this range of centralities in any of the beam energy scan experiments planned or under way.

\addtocounter{section}{1}
\section*{\large{\textsf{Acknowledgements}}}
The author thanks Dr Cate McInnes for helpful discussions.

\end{document}